\documentstyle[epsfig]{elsart}
\begin{document}

\newcommand{\mathbold}[1]{\mbox{\rm\bf #1}}
\newcommand{\mrm}[1]{\mbox{\rm #1}}
\newcommand{\N}{{\cal N}}
\newcommand{\bla}{\hspace{1cm}}
\newcommand{\be}{\begin{equation}}
\newcommand{\ee}{\end{equation}}
\newcommand{\nn}{\nonumber}
\newcommand{\bea}{\begin{eqnarray}}
\newcommand{\eea}{\end{eqnarray}}
\newcommand{\eq}[1]{eq.~(\ref{#1})}
\newcommand{\rfn}[1]{(\ref{#1})}
\newcommand{\Eq}[1]{Eq.~(\ref{#1})}
\newcommand{\gsim}{\ \rlap{\raise 2pt\hbox{$>$}}{\lower 2pt \hbox{$\sim$}}\ }
\newcommand{\lsim}{\ \rlap{\raise 2pt\hbox{$<$}}{\lower 2pt \hbox{$\sim$}}\ }
\newcommand{\ep}{\epsilon}
\newcommand{\D}{\Delta}

\newcommand{\ssu}{$SU(2)_L\times SU(2)_R\times U(1)_{B-L}\,$}
\newcommand{\sul}{$SU(2)_L$}
\newcommand{\sulu}{$SU(2)_L\times U(1)_Y$}
\newcommand{\sur}{$SU(2)_R$}
\newcommand{\thh}{\displaystyle\frac{1}{3}}
\newcommand{\matr}{\left( \begin{array}}
\newcommand{\ematr}{\end{array} \right)}
\newcommand{\g}{\gamma}
\newcommand{\lr}{{left-right symmetric model }}

\newcommand{\ea}{{ et al.}}
\newcommand{\ib}{{\it ibid.\ }}

\newcommand{\np}[1]{Nucl. Phys. {\bf #1}}
\newcommand{\pl}[1]{Phys. Lett. {\bf #1}}
\newcommand{\pr}[1]{Phys. Rev. {\bf #1}}
\newcommand{\prl}[1]{Phys. Rev. Lett. {\bf #1}}
\newcommand{\zp}[1]{Z. Phys. {\bf #1}}
\newcommand{\prep}[1]{Phys. Rep. {\bf #1}}
\newcommand{\rmp}[1]{Rev. Mod. Phys. {\bf #1}}    
\newcommand{\ijmp}[1]{Int. Jour. Mod. Phys. {\bf #1}}
\newcommand{\mpl}[1]{Mod. Phys. Lett. {\bf #1}} 
\newcommand{\ptp}[1]{Prog. Theor. Phys. {\bf #1}} 
\newcommand{\arns}[1]{Ann. Rev. Nucl. Sci. {\bf #1}}
\newcommand{\anfis}[1]{An. F\'\i s. {\bf #1}}
\makeatletter
\setlength{\clubpenalty}{10000}
\setlength{\widowpenalty}{10000}
\setlength{\displaywidowpenalty}{10000}

\vbadness = 5000
\hbadness = 5000
\tolerance= 1000
\arraycolsep 2pt

\footnotesep 14pt

\if@twoside
\oddsidemargin -17pt \evensidemargin 00pt \marginparwidth 85pt
\else \oddsidemargin 00pt \evensidemargin 00pt
\fi
\topmargin 00pt \headheight 00pt \headsep 00pt
\footheight 12pt \footskip 30pt
\textheight 232mm \textwidth 160mm

\let\@eqnsel = \hfil

\expandafter\ifx\csname mathrm\endcsname\relax\def\mathrm#1{{\rm #1}}\fi
\@ifundefined{mathrm}{\def\mathrm#1{{\rm #1}}}{\relax}

\makeatother

\unitlength1cm
\textheight 233mm

\begin{frontmatter}
\title {{\bf Low energy physics and left-right symmetry. Bounds on the
model parameters.}}

\author{ M. Czakon},
\author{ J. Gluza},
\author{M. Zra\l ek}

\address{ Department of Field Theory and Particle Physics, Institute 
of Physics, University of
Silesia, Uniwersytecka 4, PL-40-007 Katowice, Poland}

\begin{abstract}

In any gauge model with spontaneous symmetry breakdown 
the gauge boson masses and their mixings are not independent
quantities. They are interconnected through the vacuum expectation 
values (VEVs) of the Higgs sector.
We discuss the low-energy experiments, namely
electron-hadron, neutrino-hadron and neutrino-electron processes
in the frame of the Manifest Left-Right Symmetric model 
and show the impact of these dependencies
on the possible heavy gauge boson masses and mixings. 
At 90 \% C.L., we obtain $M_{Z_2} \geq 1475$ GeV  
for phenomenologically favorable models, without $W_L-W_R$ mixing and  
$M_{Z_2} \geq 1205$ GeV  in  the other extreme case when a
maximal mixing is possible.
If we consider  the \lr parameters without any constraints from the
Higgs
sector these limits get down to 410 GeV. Bounds on
the $Z_2$ mixing angle as well as the $W_2$ mass and its mixing angle
are also given.   
\end{abstract}
\end{frontmatter}

\section{Introduction}

Undoubtedly, we can say that this  decade is a further, permanent progress
in experimental high energy particle physics. Let's  modestly mention  
LEP achievements \cite{lep} as well as top discovery \cite{top}. However, 
not less
impressive results in low energy physics have been obtained. Especially much 
has been done in neutral current physics, where data coming from both
deep inelastic neutrino-hadron, neutrino-electron scattering as well as 
electron-hadron interactions have been enriched lately by exquisitely 
precise measurements
of parity nonconservation (PNC) in heavy atoms, such as cesium \cite{wood} 
and thalium \cite{thal}. The CCFR collaboration data on quark-Z boson 
couplings has also improved \cite{ccfr}.
This kind of experiments is  a nice (and not too
expensive regarding high energy collider physics) tool to 
probe the standard electroweak model (SM)
and its parameters. Moreover, searches
beyond the standard model physics using low-energy data 
complement quite well  the efforts made at high energy colliders. 
To visualize this statement we use in this work, as a representative
for `new physics', the classical Manifest Left-Right Symmetric (MLRS)
model. Its principle advantage over the SM is space inversion invariance at
high energies, implied not only by the gauge group but also by a discrete 
symmetry (replacement of the left by the right fields and {\it vice versa}). 
As a consequence the
left and right couplings $g_L$, $g_R$ are equal,
$g_L=g_R=g$, and the Yukawa matrices in the quark and lepton sectors are 
hermitian. A minimal Higgs sector with a  bidoublet $\Phi$, and 
left $\Delta_L$  and right $\Delta_R$ 
triplets is adopted \cite{lr,gun}, with the additional assumption that
only $\Phi$ and $\Delta_R$ have non-vanishing VEVs\footnote{
$$<\Phi>=\frac{1}{\sqrt{2}} \left( \matrix{\kappa_1 & 0 \cr
                       0 & \kappa_2} \right),\;
<\Delta_R>=\frac{1}{\sqrt{2}} \left( \matrix{0 & 0 \cr
                       v_R & 0} \right),\;<\Delta_L>=0.$$
}.
In this model we have four non-standard parameters, namely, additional 
gauge boson masses $M_{W_2}$, $M_{Z_2}$ and mixing angles in both charged 
and neutral gauge sectors ($\zeta$, $\phi$). We use them to 
parametrize the mentioned low-energy neutral data. 
These parameters are not 
independent of each other, as they are functions 
of the VEVs of $\Phi$ and $\Delta_{R}$. Based on this fact
we end up with two independent factors
$\gamma=M_{Z_1}^2/M_{Z_2}^2$ and $\epsilon=2 \kappa_1
\kappa_2/(\kappa_1^2+\kappa_2^2)$.  
This way, exploiting the full strength of the MLRS enabled us 
to obtain quite impressive limits on the $M_{Z_2}$
mass (much above 1 TeV).
The results are better 
than those from direct searches 
at high energy hadron colliders and comparable to those extracted from 
the LEPI data. 
 All numerics are done with the CERN  code MINUIT \cite{min}.

\section{$M_{Z_2}$ mass and low-energy neutral current experiments}

The low energy processes' momentum transfer being much smaller than the
intermediate gauge boson masses, contact four-fermion Lagrangians
can be effectively used.
Four-fermion 
neutrino-hadron $(\nu N)$,
neutrino-electron $(\nu e)$ and parity-violating
electron-hadron $(eN)$ interactions can be written in the conventional form as 
follows \cite{98}

\begin{eqnarray}
L^{\nu N} & = &- \frac{G_F}{\sqrt{2}} \bar{\nu} \gamma^{\mu}(1-\gamma_5)
\nu  \sum_{i=u,d} [ \epsilon_L(i) \bar{q}_i \gamma_{\mu}(1-\gamma_5)q_i
+ \epsilon_R(i) \bar{q}_i \gamma_{\mu}(1+\gamma_5)q_i], \nonumber \\
&&  \\
L^{\nu e}  & = &- \frac{G_F}{\sqrt{2}} \bar{\nu} \gamma^{\mu}(1-\gamma_5)
\nu \bar{e}_i \gamma_{\mu}(g_V^{\nu e}-g_A^{\nu e} \gamma_5)e, \\
L^{\nu N} & = & \frac{G_F}{\sqrt{2}} 
 \sum_{i=u,d} [ C_{1i}  \bar{e} \gamma_{\mu}\gamma_5 e  
\bar{q}_i \gamma_{\mu} q_i + C_{2i}  \bar{e} \gamma_{\mu} e  
\bar{q}_i \gamma_{\mu}\gamma_5 q_i].
\end{eqnarray}

Here, 
$\epsilon_{L,R}(i),g_{V,A}^{\nu e}, C_{ij}$ are model-dependent coefficients.
It is usual to consider only pure left-handed currents in $L^{\nu e}$
and $L^{\nu N}$. In the SM they can be derived by comparison with the neutral
current Lagrangian ( $ T^L_{3i} $ and $Q_i$ 
are the weak isospin of fermion i and its charge, respectively)
\begin{equation}
L^{SM}_{NC}=\frac{g}{2 \cos{\Theta_W}} \sum_{u,d,\nu,e} 
\bar{\Psi_i} \gamma^{\mu} \left( g_V^i-g_A^i \gamma_5 \right) \Psi_i Z_{\mu}
\end{equation}
with
\begin{eqnarray}
g_V^i & \equiv & T^L_{3i}-2Q_i \sin^2{\Theta_W}, \\
g_A^i & \equiv & T^L_{3i}. 
\end{eqnarray}
Using The SM definition
$\frac{G_F}{\sqrt{2}}=
\frac{\pi \alpha}{2 \sin^2{\Theta_W} (1-\Delta r)  M_{W_1}^2}$, we get

\begin{eqnarray}
\epsilon_{L}^{SM}(i) & = & \rho_{\nu N} ( T_{3i}-Q_i \kappa_{\nu N}
\sin^2{\Theta_W})+\lambda_{iL}, \\
\epsilon_{R}^{SM}(i) & = & \rho_{\nu N} ( -Q_i \kappa_{\nu N}
\sin^2{\Theta_W})+\lambda_{iR}, \\
C_{1i}^{SM} & = & \rho_{eq}' ( -T_{3i}+2Q_i \kappa_{eq}'
\sin^2{\Theta_W})+\lambda_{1i}, \\
C_{2i}^{SM} & = & \rho_{eq} ( -1/2+2 \kappa_{eq}
\sin^2{\Theta_W}) (2 T_{3i})+\lambda_{2i}, \\
{(g_V^{\nu e})}^{SM} & = & \rho_{\nu e} ( -1/2+ \kappa_{\nu e}
\sin^2{\Theta_W}),  \\
{(g_A^{\nu e})}^{SM} & = & \rho_{\nu e} ( -1/2). 
\end{eqnarray}

The $\rho,\lambda$ and $\kappa$ factors include the effects
of one-loop radiative corrections to the low energy processes
\cite{98}. At tree level $\rho=\kappa=1$ and $\lambda=0$.

Now let us procede to the left-right symmetric model.

The masses of the gauge bosons $(M_{Z_{1,2}},M_{W_{1,2}})$ and the 
mixing angles
$\zeta, \phi$ in charged and neutral gauge sectors are the following
($\kappa_+=\sqrt{\kappa_1^2+\kappa_2^2}$) \cite{pol,nasze}

\begin{eqnarray}
M^2_{W_{1,2}} &=& \frac{g^2}{4} \left[
\kappa^2_+ +v_R^2 \mp \sqrt{v_R^4+4 \kappa^2_1\kappa^2_2} \right], \\
M^2_{Z_{1,2}} &=& \frac{1}{4} \left\{ \left[
g^2 \kappa^2_+ +2v_R^2 \left( g^2+g'^2 \right) \right] \right. \nonumber \\
& \mp & \left.  \sqrt{
\left[ g^2 \kappa^2_+ +2v_R^2 \left( g^2+g'^2 \right) \right]^2- 4g^2
\left( g^2+2g'^2 \right) \kappa^2_+ v_R^2 } \right\}, \\
\tan{2\xi}&=&-\frac{2\kappa_1 \kappa_2}{v_R^2}, \\
\sin{2\phi}&=&
-\frac{g^2 \kappa^2_+ \sqrt{\cos{2\Theta_W}}}{2\cos^2{\Theta_W}
\left( M^2_{Z_2}-M^2_{Z_1} \right)}.
\end{eqnarray}

Obviously these  are functions of the three VEVs $v_R$, $\kappa_1,\kappa_2$.
Certainly, $M_{W(Z)_1} <  M_{W(Z)_2}$, so $\kappa_+^2,\kappa_1 \kappa_2
 << v_R^2$ and
we expand the above formulas leaving terms up to $O\left( 
(\frac{\kappa_+}{v_R})^2,
(\frac{\kappa_1 \kappa_2}{v_R^2}) \right)$
(we will comment on this approximation at the end of the Chapter).

The result can be cast in the form
$\left( \beta=\frac{M_{W_1}^2}{M_{W_2}^2}  \right) $ 
\cite{pol,nasze,mas}

\begin{eqnarray}
\gamma \equiv \frac{M_{Z_1}^2}{M_{Z_2}^2} & = & \frac{\cos{2 \Theta_W}}
{2 \cos^4{\Theta_W}} \beta , \\
\zeta & = & -\epsilon \beta,  \\
\phi & = & -\frac{(\cos{2\Theta_W})^{3/2}}{2 \cos^4{\Theta_W}} \beta , \\
\mbox{\rm and} && \nonumber \\
\rho_{LR} \equiv \frac{M_{W_1}^2}{M_{Z_1}^2 \cos^2{\Theta_W}}  & = &
1+\left[-\epsilon^2+\frac{1}{2} (1-\tan^2{\Theta_W})^2 \right] \beta ,
\end{eqnarray}
where
\begin{equation}
\epsilon = \frac{2 \kappa_1 \kappa_2}{\kappa_1^2+\kappa_2^2}, \;\;\;
0 \leq \epsilon \leq 1.
\end{equation}

In the MLRS model the neutral current interaction for any fermions can 
be written 

\begin{equation}
L_{NC}=\frac{e}{2\sin{\Theta_W}\cos{\Theta_W}}
\sum\limits_{i=up,\;down,l,\nu }\sum\limits_{j=1,2} \bar{\psi}_i\gamma^\mu
\left[
A^{ji}_L \Omega_L^{i}P_L + A^{ji}_R \Omega_R^{i} P_R
\right]
\psi_i Z_{j\mu}.
\end{equation}

The couplings $A^{1,2;\;i}_{L,R}$ are given by

\begin{eqnarray}
A_L^{1i} & = & \cos{\phi}\;g^i_L+\sin{\phi}\;g_L^{\prime i},\\
A_R^{1i} & = & \cos{\phi}\;g^i_R+\sin{\phi}\;g_R^{\prime i},\\
A_L^{2i} & = & \sin{\phi}\;g^i_L-\cos{\phi}\;g_L^{\prime i},\\
A_R^{2i} & = & \sin{\phi}\;g^i_R-\cos{\phi}\;g_R^{\prime i},
\end{eqnarray}

where

\begin{eqnarray}
g^i_L & = & 2 T^L_{3i}-2Q_i\sin^2\Theta_W , \\
g_L^{\prime i} & = & \frac{2\sin^2\Theta_W }{\sqrt{\cos 2\Theta_W }}
\left( Q_i-T^L_{3i}\right), \\
g^i_R & = & -2 Q_i \sin^2 \Theta_W , \\
g_R^{\prime i} & = & \frac{2}{\sqrt{\cos 2\Theta_W }}
\left( Q_i \sin^2 \Theta_W - T^R_{3i}\cos^2 \Theta_W \right).
\end{eqnarray}

$\Omega_{L,R}$ are analogous to Cabbibo-Kobayashi mixing matrices in the
charged sector and are the identity matrices for charged fermions
$$\Omega_{L,R}^{i}=I\;\;\; \mbox{\rm for}\;\; i=u,d,l.$$
To have a link with the model independent Lagrangians (Eq.(1-3))
we now assume that only 
pure left-handed neutrinos play a role in neutral low energy 
physics\footnote{If it was not true then we would certainly have had an 
indication on the long standing Dirac-Majorana neutrino nature problem
\cite{zral}.}.
Then $\Omega_L^{\nu} \simeq I, \Omega_R^{\nu} \simeq 0$.

We can now, quite analogously to the SM case, find low energy 
LR model coefficients

\begin{eqnarray}
\epsilon_{L,R}^{LR}(i) & = & \Lambda ( A_L^{1 \nu}
A_{L,R}^{1i}+ \gamma A_L^{2\nu}A_{L,R}^{2i} ), \\
C_{1i}^{LR} & = & \Lambda(g_A^{1l} g_V^{1i} + \gamma g_A^{2l} g_V^{2i} ), \\
C_{2i}^{LR} & = & \Lambda(g_V^{1l} g_A^{1i} + \gamma g_V^{2l} g_A^{2i} ), \\
{(g_V^{\nu e})}^{LR} & = & \Lambda(A^{1\nu}_L g_V^{1l} 
+ \gamma A^{2 \nu}_L g_V^{2l} ), \\
{(g_A^{\nu e})}^{LR} & = & \Lambda(A^{1\nu}_L g_A^{1l} 
+ \gamma A^{2 \nu}_L g_A^{2l} ), 
\end{eqnarray}
where
\begin{eqnarray}
g_{V,A}^i & = & \frac{1}{2} \left( A_L^{i \nu} \pm A_R^{i \nu} \right), \\
\Lambda &=& \frac{\rho_{LR}}{ \left( \cos^2{\zeta}+\beta \sin^2{\zeta}
\right)} .
\end{eqnarray}
The $\Lambda$ factor is connected with the L-R definition of the $G_F$ 
constant. If we assume the only natural situation when right-handed neutrinos
are too heavy to be directly produced in the muon decay and the light are 
left-handed (negligible right-handed admixture) then
the $G_F$ definition follows \cite{cz}

\begin{equation}
\frac{G_F}{\sqrt{2}}=
\frac{\pi \alpha}{2 \sin^2{\Theta_W} M_{W_1}^2 (1-\Delta r)} 
\left( \cos^2{\zeta}+\beta \sin^2{\zeta} \right) .
\end{equation}

The parameters of Eqs. (31-35) are
`bare',  reproducing the 'bare' SM couplings for
$\phi=\gamma=0$. 
We improve them by adding the SM corrections 
(Eqs.(7)-(12)), so the data is fitted with

\begin{eqnarray}
\epsilon_{L,R}^{LR}(i) & = & \Lambda \left[ A_L^{1 \nu} \left
 ( \cos{\phi}
\epsilon_{L,R}^{SM}(i) + \frac{1}{2}\sin{\phi}
g'^{i}_{L,R} \right) + \frac{1}{2} \gamma A_L^{2 \nu} A_{L,R}^{2i} \right], \\
C_{1i}^{LR} & = &   \Lambda \left[ \cos{2 \phi}- \sin{2 \phi} 
\frac{\sin^2{\Theta_W}}{\sqrt{\cos{2 \Theta_W}}} \right]
\left[ C_{1i}^{SM}- \gamma \left( -T_{3i}+2Q_i \sin^2{\Theta_W} \right) 
\right], \\
C_{2i}^{LR} & = &   \Lambda \left[ \cos{2 \phi}- \sin{2 \phi} 
\frac{\sin^2{\Theta_W}}{\sqrt{\cos{2 \Theta_W}}} \right]
\left[ C_{2i}^{SM}- \gamma \left( -\frac{1}{2}+2 \sin^2{\Theta_W} \right) 
(2T_{3i}) \right] \nonumber \\
&& \\
{(g_V^{\nu e})}^{LR} & = &   \Lambda    \left[ 
A_L^{1 \nu} \left( \cos{\phi}-\frac{\sin{\phi}}{\sqrt{\cos{2\Theta_W}}}
\right)
{(g_V^{\nu e})}^{SM} \right. \nonumber \\
&+& \left. \gamma A_L^{2 \nu} \left( \sin{\phi}+\frac{\cos{\phi}}
{\sqrt{\cos{2\Theta_W}}} \right)  \left( -\frac{1}{2}+2 \sin^2{\Theta_W} 
 \right) \right] ,\\
{(g_A^{\nu e})}^{LR} & = &   \Lambda
 \left[ A_L^{1 \nu} ( \cos{\phi}+\sin{\phi}\sqrt{\cos{2\Theta_W}})
{(g_A^{\nu e})}^{SM} \right. \nonumber \\ 
&+& \left. \gamma A_L^{2 \nu} \left( \sin{\phi}-\cos{\phi}
\sqrt{\cos{2\Theta_W}} \right)  \left( -\frac{1}{2} \right) \right]. 
\end{eqnarray}

In Table 1 we show the 1998 data \cite{98} for all the couplings
that are used. 
As the Standard Model one-loop corrections to these 
theoretical formulas (Eqs.(7)-(12)) have been  calculated in the 
$\overline{MS}$ scheme, we take the value of $\sin^2{\Theta_W}$ in the same
scheme
$\sin^2{\Theta_W} \equiv  \hat{s}_Z^2 =0.23124 \pm 0.00017$ \cite{98}.
In Fig.1 we show the 90 \% C.L. allowed region for 
$\gamma-\phi$ parameters. The dotted line shows the results for the
data from Table 1. The solid line follows from supplementing the
previous with an 
 additional parameter which measures the neutral to charged current cross 
section ratio in neutrino scattering off nuclei
and is given by the CCFR collaboration  \cite{ccfr}.
This parameter in the frame of our model should be defined in the following 
way 
\be
\kappa^2=  1.7897 g_L^2+1.1479 g_R^2-0.0916 \delta_L^2-0.0782 \delta_R^2
\ee

where
\begin{eqnarray}
g_{L,R}^2& = & \left( \epsilon_{L,R}^{LR}(u) \right)^2+ \left( 
             \epsilon_{L,R}^{LR}(d) \right)^2, \\
\delta_{L,R}^2& = & \left( \epsilon_{L,R}^{LR}(u) \right)^2- \left( 
             \epsilon_{L,R}^{LR}(d) \right)^2. 
\end{eqnarray}

The CCFR collaboration has found it to be 

\be
\kappa^2=0.5820 \pm 0.0041.
\ee

We can see that $\kappa^2$ does not change the predictions for the $\gamma$ 
parameter. Using Eq.(17) we get $M_{Z_2} \geq$ 410 GeV. This result is
not substantially different from other analyses \cite{pol,chay}. 

\begin{table}[h]
\centering
\begin{tabular}{c c c}
\cline{1-3}
 &  \ Experimental Value & \ Correlations   \\
\cline{1-3}
$\epsilon_L(u)$ &    $ 0.328 \pm 0.0016$  &    \\
$\epsilon_L(d)$ &    $ -0.440 \pm 0.011$  &  non-  \\
$\epsilon_R(u)$ &    $ -0.179 \pm 0.0013$  & Gaussian   \\
$\epsilon_R(d)$ &    $ -0.027 \matrix{ +0.077 \cr -0.048}$  &    \\
\cline{1-3}
$g_L^2$ & $0.3009 \pm 0.0028$  &   \\
$g_R^2$ & $0.0328 \pm 0.003$  &   \\
$\Theta_L$ & $2.50 \pm 0.035$  &  small \\
$\Theta_R $ & $4.56 \matrix{ +0.42 \cr -0.27}$  &   \\
\cline{1-3}
$g_V^{\nu e}$ & $-0.041 \pm 0.015$  & $-0.04$   \\
$g_A^{\nu e}$ & $-0.507 \pm 0.014$  &   \\
\cline{1-3}
$C_{1u}$ & $-0.216 \pm 0.046$ & $-0.997$ $-0.78$ \\ 
$C_{1d}$ & $0.361 \pm 0.041$ &          0.78 \\ 
$C_{2u}-\frac{1}{2}C_{2d}$ & $-0.03 \pm 0.12$ & \\
\cline{1-3} 
\end{tabular}
\caption{\footnotesize{
Data used for the neutral data analysis \cite{98}. Appropriate formulas are
given in the text. $g_{L,R}^2=\epsilon^2_{L,R}(u)-\epsilon^2_{L,R}(d)$,
$\tan{\Theta_{L,R}}=\frac{\epsilon_{L,R}(u)}{\epsilon_{L,R}(d)}$}}
\end{table}

Until now the left-right observables  
$\beta,\gamma,\zeta,\phi$ have been  treated as independent, i.e.
we do not take into account the relations (13)-(16)  
which reflect the fact that  the VEVs link them to one another. 
When we use these relations the situation changes substantially (Fig2).
We have chosen as independent two phenomenologically handful  
parameters: $\epsilon$ and $\gamma$.
Different $\epsilon$'s describe left-right models with different bidoublet
VEVs $\kappa_1,\kappa_2$. We know from
phenomenological considerations that the reduction of FCNC favors 
left-right models with 
$\epsilon \simeq 0$ \cite{gun} (ellipses denoted with (b)). 
Then (see Eq.(18)) there is no $W_L-W_R$ mixing.
However, to make this possibility open we also 
show the results for left-right models with $\epsilon \simeq 1$ (
ellipses denoted with (a)). 
The dotted ellipses are obtained when all of the data from Table 1 is taken 
into account.
The solid ones correspond to the inclusion of the 
$\kappa^2$ Eq.(47) parameter.
Fig.2 shows that the MLRS  relations among the 
fitted parameters $\beta,\gamma,\phi,\zeta$,  and the CCFR data
make it possible to shrink considerably the allowed
space for the $\gamma$ factor. From Eqs. (17)-(20) it is possible to find 
limits on the rest of the
left-right parameters:

\begin{eqnarray*}
\left. \matrix{ 
M_{Z_2} \geq 1475\;  \mbox{\rm GeV} \cr 
M_{W_2} \geq 875\; \mbox{\rm GeV}  \cr
| \phi | \leq 0.0028\; \mbox{\rm rad}} \right\}
&& \mbox{\rm for models without }\; W_L-W_R\; \mbox{\rm mixing}, \\
&& \\
 \left. \matrix{ 
M_{Z_2} \geq 1205\; \mbox{\rm GeV} \cr 
M_{W_2} \geq 715\; \mbox{\rm GeV}  \cr 
| \zeta | \leq 0.013\; \mbox{\rm rad} \cr
| \phi | \leq 0.0042\; \mbox{\rm rad}
} \right\}
&& \mbox{\rm for models with possible}
W_L-W_R \; \mbox{\rm mixing}  \; (\epsilon \neq 0)  
\end{eqnarray*}

These results are comparable with previous LEPI analyses ($M_{Z_2}
\geq 0.8 \div 1.5$ TeV) \cite{lep1}
and better than that which follow from direct searches for additional 
gauge bosons  in hadron
colliders ($M_{Z_2} \geq 630$ GeV) \cite{hc}. 
Finally, let us comment on the approximation made
in Eqs.(17)-(20). Our fitted observables (Eq.(36)-(41)) 
are functions of $\beta,\gamma$
(so $M_{W_2},M_{Z_2}$) and mixing angles $\zeta,\phi$.
Taking into account exact formulas (13)-(16) for these quantities we
have checked that, at 90 \% C.L., quantities of the form
$(\kappa_+/v_R)^2$ and $\kappa_1 \kappa_2/v_R^2$ do not exceed
at the worst case 0.04 so, when neglecting squares of them, 
relations (17)-(20) are quite reliable.

\section{Conclusions}

We point out the importance of examining non-standard
models using relations
among physical parameters such as masses and mixings that follow
from the Higgs sector. 
In our analysis we used the most up to date low energy experimental data,
including the CCFR $\kappa^2$. 
Thanks to these we obtained
new limits on \lr parameters that are 
comparable with those from high energy physics. 

\begin{ack}
This work was supported by Polish Committee for Scientific Research under 
Grants Nos. 2P03B08414 and 2P03B04215. 
J.G. appreciates also the support of the Humboldt-Stiftung.
\end{ack}

\newpage 

\begin{figure}
\epsfig{figure=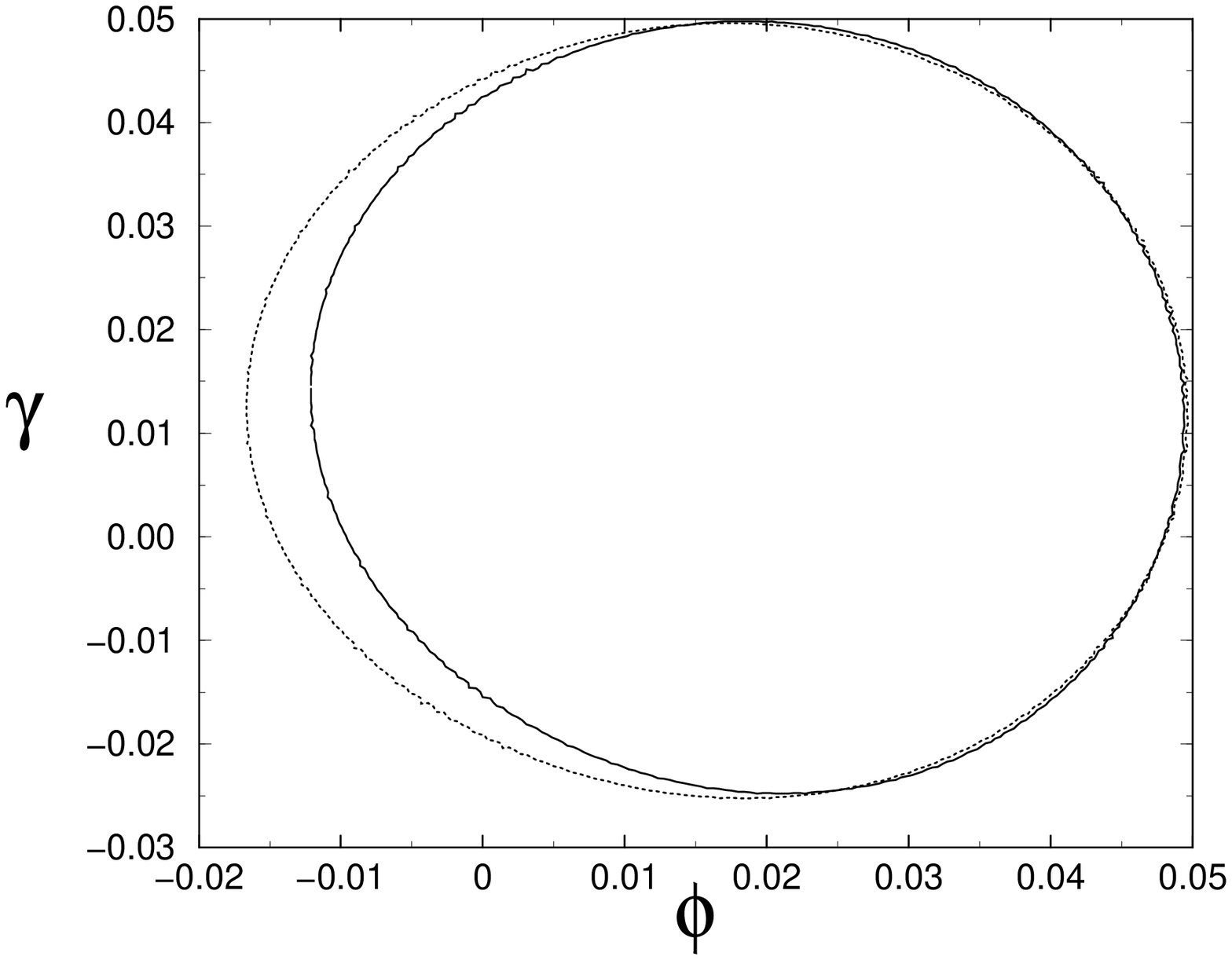, height=5 in}
\caption{90 \% C.L. region for allowed $\gamma-\phi$ parameters.
Dashed line describes result when data of Table 1 are taken into account.
Solid line takes into account additional data given in Eq.(47).}
\end{figure}

\begin{figure}
\epsfig{figure=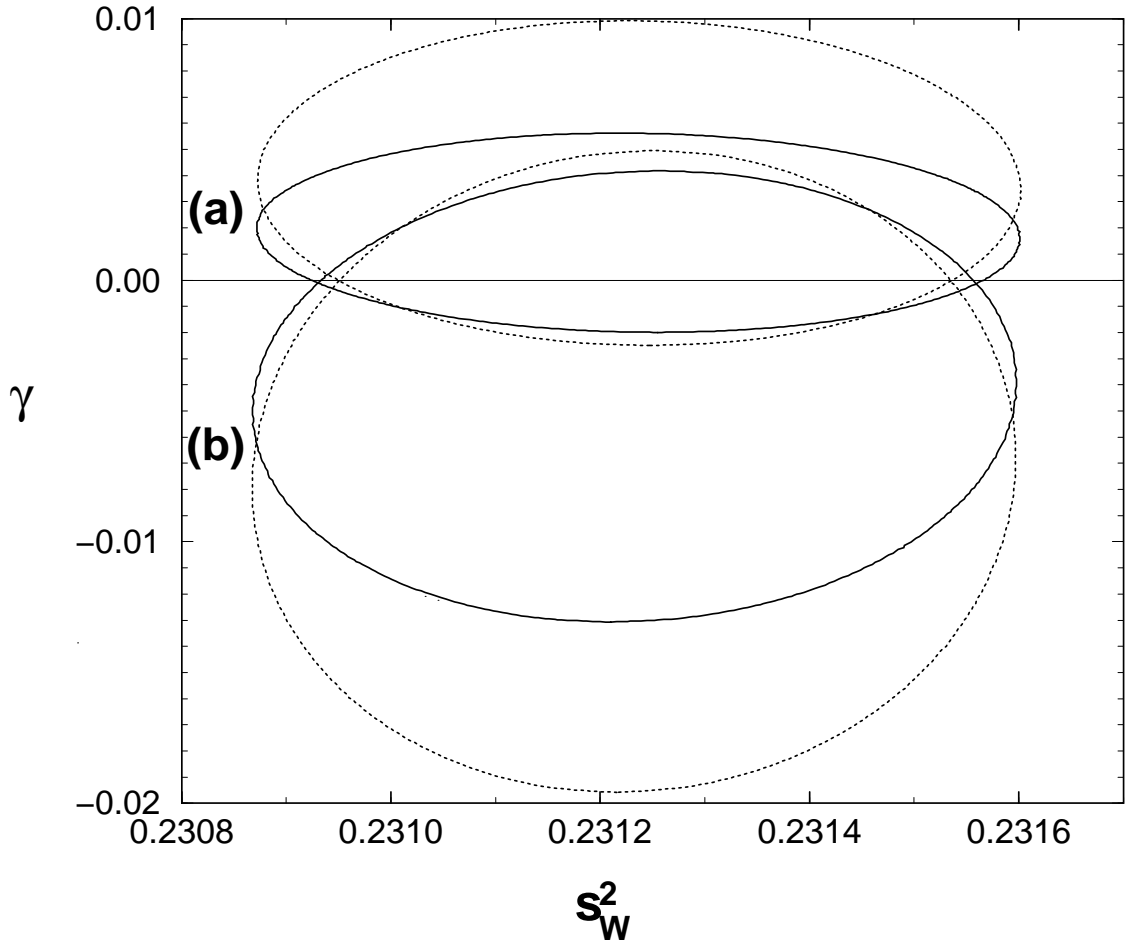, height=5 in} 
\caption{90 \% C.L. region for the allowed $\gamma-\sin^2{\Theta_W}$
parameters when relations Eqs.(17-21) are taken into account. 
Two upper ellipses (a) realize models with $\epsilon=1$
(possible $W_L-W_R$ mixing). Two lower ellipses (b) give results for 
$\epsilon=0$
(no $W_L-W_R$ mixing). The dashed line describes the result when the data 
of Table 1 
is taken into account.
The solid line takes into account the additional data given in Eq.(47).}

\end{figure}


\end{document}